\documentclass[pre,twocolumn,twoside,byrevtex,superscriptaddress,floatfix]{revtex4-1}

\usepackage{settings}
\usepackage{float}
\usepackage[margin=1in,columnsep=.8cm]{geometry}
\usepackage{multirow}
\usepackage{ragged2e}
\usepackage[rightcaption]{sidecap}    
\usepackage{array}
\usepackage{longtable}

\definecolor{lightgray}{gray}{0.9}
\setboolean{twocolswitch}{True}
\begin{document}

\title{\protect The incel lexicon: \\
Deciphering the emergent cryptolect 
of a 
global misogynistic community



}
\author{
\firstname{Kelly}
\surname{Gothard}
}
\affiliation{
  Computational Story Lab,
  Vermont Complex Systems Center,
  MassMutual Center of Excellence for Complex Systems and Data Science,
  Vermont Advanced Computing Core,
  The University of Vermont,
  Burlington, VT 05401.
}


\author{
\firstname{David Rushing}
\surname{Dewhurst}
}
\affiliation{
  Charles River Analytics, 
  625 Mount Auburn Street, 
  Cambridge, 
  MA 02138.
}
\author{
\firstname{Joshua R.}
\surname{Minot}
}
\affiliation{
  Computational Story Lab,
  Vermont Complex Systems Center,
  MassMutual Center of Excellence for Complex Systems and Data Science,
  Vermont Advanced Computing Core,
  The University of Vermont,
  Burlington, VT 05401.
}
\author{
\firstname{Jane Lydia}
\surname{Adams}
}
\affiliation{
  Computational Story Lab,
  Vermont Complex Systems Center,
  MassMutual Center of Excellence for Complex Systems and Data Science,
  Vermont Advanced Computing Core,
  The University of Vermont,
  Burlington, VT 05401.
}
\author{
\firstname{Christopher M.}
\surname{Danforth}
}
\affiliation{
  Computational Story Lab,
  Vermont Complex Systems Center,
  MassMutual Center of Excellence for Complex Systems and Data Science,
  Vermont Advanced Computing Core,
  The University of Vermont,
  Burlington, VT 05401.
}
\affiliation{
Department of Mathematics \& Statistics, University of Vermont, Burlington, VT 05401.}
\affiliation{
Department of Computer Science, University of Vermont, Burlington, VT 05401.}

\author{
\firstname{Peter Sheridan}
\surname{Dodds}
}
\affiliation{
  Computational Story Lab,
  Vermont Complex Systems Center,
  MassMutual Center of Excellence for Complex Systems and Data Science,
  Vermont Advanced Computing Core,
  The University of Vermont,
  Burlington, VT 05401.
}
\affiliation{
Department of Mathematics \& Statistics, University of Vermont, Burlington, VT 05401.}
\affiliation{
Department of Computer Science, University of Vermont, Burlington, VT 05401.}

\date{\today}

\begin{abstract}
  \protect
  Evolving out of a gender-neutral framing of
an involuntary celibate identity,
the concept of `incels' 
has come to refer to an online community of men 
who bear antipathy towards themselves, women, and society-at-large
for their perceived inability to find and maintain sexual relationships.
By exploring incel language use on Reddit, 
a global online message board, 
we contextualize the incel community's 
online expressions of misogyny 
and real-world acts of violence perpetrated against women.
After assembling around three million comments from
incel-themed Reddit channels, 
we analyze the temporal dynamics of a 
data driven rank ordering of the glossary of 
phrases belonging to an emergent incel lexicon. 
Our study reveals the generation and normalization
of an extensive coded misogynist vocabulary 
in service of the group's identity.

\end{abstract}



\maketitle


\section{Introduction}\label{sec:introduction} 

Incels are self-identified members of a global, online subculture whose members subscribe to resentful misogynist views on women as the result of a perceived unfulfilled entitlement to love and sex.  
Incels publicly participate in discussions on Reddit, 4chan, and other platforms where pseudo-anonymity can be preserved. 
Online incel discussions are in part encoded by slang terms 
as well as memes which express misogyny through humor,
sufficient to generate an incel cryptolect.

\begin{figure*}[tb!]
\centering
\includegraphics[width=\textwidth, height=4.5cm]{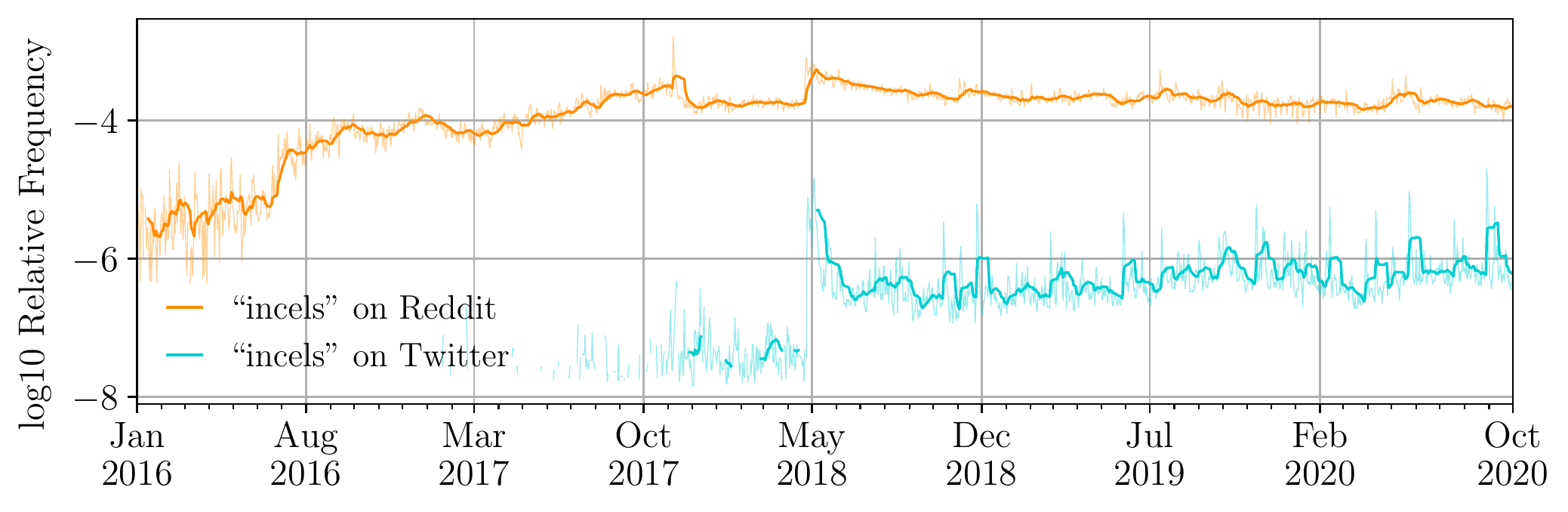}
\caption{\textbf{Relative frequency of `incels' on Reddit and Twitter}. Appearances of the word `incels' on Twitter and Reddit are enumerated and visualized using data from the StoryWrangler API \cite{alshaabi2020storywrangler} and Pushshift Reddit API \cite{baumgartner2020pushshift}. The StoryWrangler API makes publicly available the top one million most popular phrases daily, rendered here as a four year relative frequency time series (including retweets) of messages containing `incels'. The term `incels' does not maintain a steady presence in the top one million most frequent words on Twitter until its relative frequency spiked following the April 23, 2018 Toronto van attack.  On Reddit, `incels' sees growth in July 2016, a small dip following the ban of r/Incels, and a spike on the day of the Toronto van attack.  This attack is associated with increased conversation about incels on both Reddit and Twitter. The roughly two order of magnitude disparity in relative frequency is due to underlying differences in accounts and demographics of individuals reflected by the two corpora.}
\label{fig:twitter_reddit}
\end{figure*}


Within the last decade, incels have gained notoriety for their toxic contributions to online communities and for their association with real-world acts of mass violence. On May 23, 2014, 22 year old Elliot Rodger killed 6 people and injured 14 others shortly after sending a written manifesto to people he knew, as well as uploading a video to YouTube that detailed his loathing for women and anger towards society. Later perpetrators have cited Elliot Rodger as an inspiration for their attacks \cite{wiki:islavista}.  For example, on April 23, 2018, Alek Minassian drove a van into a Toronto crowd, killing 10 and injuring 16, an hour after posting the following text to Facebook:

\begin{quote}
Private (Recruit) Minassian Infantry 00010, wishing to speak to Sgt 4chan please. C23249161. The Incel Rebellion has already begun! We will overthrow all the Chads and Stacys! All hail the Supreme Gentleman Elliot Rodger! \cite{Madhani.2018}
\end{quote}

A number of other attacks have been associated with the incel community, 
such as the Tallahassee yoga studio shooting, during which Scott Beierle shot and killed two women shortly after referencing Elliot Rodger in videos online \cite{wiki:tallahassee}.  
These attacks brought the threat of incels to the forefront of US media attention, as well as that of the US Air Force, which held a briefing in 2019 to discuss the increasing national threat of incel attacks \cite{airforce.2019}.  The Southern Poverty Law Center (SPLC) found that the incel community not only praised attackers who identified as incels, but also praised the Las Vegas shooter for killing `normies', despite his lack of association with the incel community \cite{janik2018}.  
The incel community praises mass acts of violence and does so using a unique lexicon that has gradually crept into popular culture.

Inceldom is part of a larger misogynist ecosystem, called the `Manosphere'
\cite{Ging:2019,marwick2018drinking}.  
In addition to incels, the Manosphere is comprised of Men's Rights Activists, Men Going Their Own Way (MGTOW), and Pick-Up Artists (PUAs).  
Each of these groups subscribe to the same underlying philosophy, referred to as the `red pill' \cite{Ging:2019}.  When an individual has `taken the red pill', they have enlightened themselves to a reality in which women wield feminism as a weapon against men, depriving them of sex and love. The phrase originally appeared in the film `The Matrix' under a different context \cite{thematrix}. PUAs seek to regain sexual power by taking advantage of women; MGTOW members voluntarily reject relationships with women altogether; 
and incels commiserate and express anger over their lack of sexual activity.  
Here, we narrow our focus to the incel community on Reddit.



Incel communities exist on a variety of platforms including Reddit, 4chan, and other independent websites like incels.co \cite{manivannan2013fcj, nagle2015investigation}.  Reddit is a platform that hosts subreddits, topic-specific forums that are created and moderated by users. 
The no longer active subreddit r/Incels was 
considered by the community to be one of the first incel message boards to exist on Reddit \cite{inceltimeline:2020}.  As part of a larger effort to limit the impact of violent content on the site, Reddit banned r/Incels on November 7, 2017 for inciting violence against women \cite{Hauser.2017}.

As is commonly witnessed when communities promoting hate speech are disbanded, new gathering places emerge. 
`Interconnected hate clusters form global ‘hate highways’ that---assisted by collective online adaptations---cross social media platforms, sometimes using ‘back doors’ even after being banned, as well as jumping between countries, continents, and languages' \cite{Johnson-Nature}.

Shortly after r/Incels was banned, r/Braincels gained popularity and the Toronto van attack took place. In the following months, Reddit quarantined r/Braincels twice, for a short period on September 25, 2018, and from January 2019 through May 2019. When a subreddit is quarantined, posts and comments do not appear on Reddit's popular public subreddits like r/all, and the subreddit is not included in promoted subreddits.

The r/Braincels subreddit was banned permanently on September 30, 2019, shortly after the US Army sent out a memo to service members regarding the growing threat of violence from incels, triggered by FBI intelligence on potentially violent incel activity associated with the theatrical release of the film \textit{The Joker (2019)} \cite{Cameron.2019}. In response, various other replacement subreddits including r/Shortcels showed early signs of popularity, but many were banned after only a few weeks \cite{Futrelle.2019}.  Incel subreddits still exist on Reddit and are banned only when they violate Reddit's specific community standards.  Many subreddits stay in existence by pushing the envelope of what is acceptable on Reddit and what is not.

In addition to Reddit, the incel community is discussed on other popular social media sites including Twitter. Fig.~\ref{fig:twitter_reddit} illustrates the prevalence of the word `incels' on both Reddit and Twitter, and the quick adoption of the term on Twitter following the Toronto van attack in April 2018, an event that catapulted awareness of the incel community to national prominence.

\begin{figure}[t]
\centering
\includegraphics[width=\linewidth]{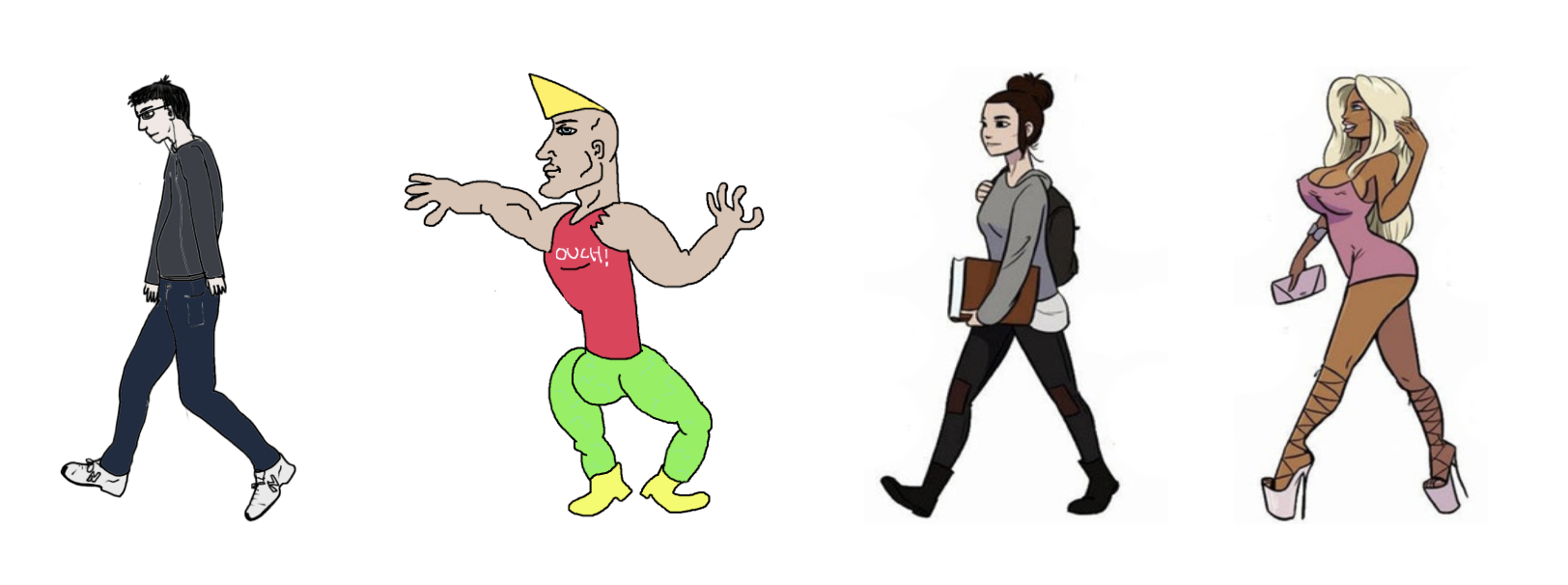}
\caption{\textbf{Popular incel Memes (left to right): Virgin, Chad, Becky and Stacy.} The `Virgin' represents the incel, and is depicted with his head down, a thin stature and pale skin \cite{virginchad:2019}. He is characterized as inferior to a `Chad', who is brutish and unintelligent \cite{virginchad:2019}.  `Becky' represents an undesirable woman, who is educated and outspoken \cite{beckystacy:2018}.  Lastly, `Stacy' is the female counterpart to Chad.  She is represented as promiscuous and shallow \cite{beckystacy:2018}.}
\label{fig:meme}
\end{figure}

Incel language and memes are associated with acts of misogynist violence, and as they become popular, they spread such misogyny. Alek Minassian's previously referenced quote includes `Chads', and `Stacys', among other phrases associated with Incels.  Fig.~\ref{fig:meme} describes the `Chad', `Stacy', `Virgin', and `Becky' memes which categorize men and women by their sexual attractiveness. In their recent study of online rhetoric, Ging \etal\ found that `rape-glish' and `gendered e-bile' have infiltrated Urban Dictionary, indicating a broader integration of misogynistic language into popular culture \cite{ging2020neologising}.  Because Urban Dictionary is crowd-sourced, Ging 
\etal\ postulate that its contents are a reflection of popular language trends, and could therefore determine if misogynistic language was highly prevalent by measuring its presence on Urban Dictionary using machine learning methods. Emma Jane also illustrated the prevalence of misogynistic language online by creating and Random Rape Threat Generator, highlighting how popular and formulaic certain phrases have become \cite{jane2018rapeglish}.

Reddit and 4chan are online spaces where many users virtually self-identify as living on the margins of society. To compensate for these feelings, male participants in these communities often treat women as intruders, and so female redditors are often met with more harassment and degradation than their male counterparts \cite{doi:10.1080/14680777.2016.1120490, gingsiaperaspecial, massanari2017gamergate, banet2016masculinitysofragile}.  The humorous nature of memes provides a blanket of social safety to cover up the racism/sexism---making it difficult to distinguish satire from genuine bigotry, as illustrated through racist memes on 4chan \cite{milner2013fcj}.  The `Virgin v.s. Chad' meme is a notable example of incel lexicon proliferation.  According to KnowYourMeme.com, the `Virgin v.s. Chad' meme originated on 4chan and was first seen on Reddit in June 2017 on r/4chan and r/justneckbeardthings \cite{knowyourmeme:2017}.  The meme depicts caricatures of the virgin incel and of the alpha-male Chad, a made-up character that embodies stereotypical masculinity.  The meme is popularly used to compare any two things, and to show that one is very clearly inferior.  There is also a female variant of the meme: the feminist and dowdy Becky versus the shallow and hyper-sexual Stacy. The existence of a language unique to misogyny gives perceived power and legitimacy to those who use it, and relies on humor to make it palatable.

The `Virgin vs Chad' meme is meant to be self-referential to incels, which gives it an even greater blanket of protection.  However, it reinforces negative stereotypes about true masculinity and rape culture, and quickly gives way to a blatantly sexist version of the meme about women.  In her study of BlueSky MUD, an `interactive, text-only online forum...based on the original multiperson networked Dungeons \& Dragons–type game called MUD', Lori Kendall finds that misogyny is expressed through seemingly-harmless humor. The geeky and socially uninclined participants make jokes about their own ineptitude for seducing women `regarding their non-hegemonic masculinity, but women are the ultimate butts of the joke' \cite{kendall2002hanging}.  This particularly popular incel meme disguises itself as self-deprecating humor, when in actuality, it is a vehicle for normalizing sexism.  The popularity of the `Virgin vs Chad' meme, and by extension, terms in the incel lexicon, reflects a broader public appetite for consuming and producing sexist content \cite{phillips2012we, knobel2007new, shifman2013memes}.

In this study, we aim to dismantle the ambiguity and humor which masks the toxic context in which terms such as `Chad' and `virgin' exist by empirically identifying a contextualized incel lexicon.  We present a comparison of over 3.5 million comments from three popular incel subreddits---r/Incels, r/Braincels, and r/Shortcels---to 2 million randomly selected Reddit comments. To reveal the incel lexicon without introducing researcher bias, or the bias of predetermined topics or anchor words, we use a rank-based divergence measure coupled with a birds-eye system comparison visualization method that illuminates important terms at all rankings, rather than just high frequency terms.  We then explore the timeseries of each word in the empirically derived incel lexicon to determine whether those words have a stable presence.  To build context around the terms in the incel lexicon, we use rank-turbulence divergence to determine the most important bigrams and trigrams.  By identifying and contextualizing an empirically derived incel lexicon, we summarize key terms and themes in incel discourse.

In Section II, we describe the Reddit comments data used to study incel language and activity patterns, as well as the method by which we compare rank distributions of words in incel vs random comments.  In Section III, we present and discuss the results of our analysis.

\section{Methods}\label{sec:methods}
\subsection{Data}

We retrieved Reddit comments from the open-source Pushshift Reddit dataset and the Python Reddit API Wrapper (PRAW) \cite{baumgartner2020pushshift, boe2016python}.  Data from the Pushshift Reddit dataset is comprised of monthly snapshots of all comments on Reddit, which allows us to access comments from banned subreddits that would not be accessible through the Reddit API.  We retrieved comments for the random corpus using the PRAW package instead of the Pushshift API because PRAW has features for randomly selecting subreddits, posts and comments, whereas the Pushshift Reddit API does not.  We compiled the random corpus by randomly selecting a subreddit, then randomly selecting a post from that subreddit, and finally randomly selecting a comment from that post thread, via PRAW.

\begin{table*}[t!]
\centering
\begin{tabular}{l|r|r|r}
\hline
 Subreddit & Total Unique Users & Total Comments & Comments/User\\
\hline
r/Incels & 36,301 & 881,118 & 24.27\\
r/Braincels & 56,698 & 2,486,655 & 43.86\\
r/ShortCels & 9,264 & 214,313 & 23.13\\
\hline
\textbf{Incel Corpus Total} & \textbf{94,704} &  \textbf{3,570,548} & \textbf{37.70}\\
\hline
Random & 925,310 & 2,082,274 & 2.25\\
\hline
r/TwoXChromosomes & 543,534 & 3,364,567 & 6.19 \\
\hline
r/liberal & 24,717 & 152,822 & 6.18 \\
\hline
r/conservatives & 37,304 & 310,100 & 8.31 \\
\hline
\end{tabular}
\caption{\textbf{Total comments, users, and comments per user in the incel and random comments datasets.}  All available comments from r/Incels, r/Braincels, r/ShortCels, r/TwoXChromosomes, r/liberal, and r/conservatives were collected from the Pushshift Reddit dataset.  The random set of comments were retrieved from the Reddit API using PRAW.}
\label{table:data_summary}
\end{table*}

The total number of comments and users per corpus are listed in Table \ref{table:data_summary}.  We chose r/Incels, r/Braincels, and r/Shortcels to populate the incel corpus because they were found to be the most popular subreddits, according to both an incel wiki and journalist David Futrelle from \textit{We Hunted the Mammoth} \cite{inceltimeline:2020, Futrelle.2019}. This dataset does not capture the entire corpus of communication within the incel community, but by selecting very popular subreddits, we gain knowledge about common incel language use.  We used comments from the popular feminist subreddit r/TwoXChromosomes to control for topics related to gender and sexuality \cite{twoxchromosomes}.  Additionally, we chose r/liberal and r/conservatives to represent common ideologies so that we may compare levels of user activity in these subreddits to user activity levels in the incel and feminist subreddits.

We cleaned text data from comments by removing entries with deleted user names, deleted text bodies, and missing UTC timestamps.  Additionally, we removed comments authored by user ``AutoModerator" because this user name is associated with a programmable bot created by Reddit that can be programmed by moderators of subreddits to autonomously make moderation comments when other users comment.  The remaining entry totals after cleaning are recorded in Table \ref{table:data_summary}. We cleaned the text body of each comment by removing punctuation, 1-grams that contain ``http" as a method for removing links, and HTML artifacts such as ``\&gt", ``x200b", and ``\&amp".

\begin{figure}[t]
\centering
\includegraphics[width=\columnwidth]{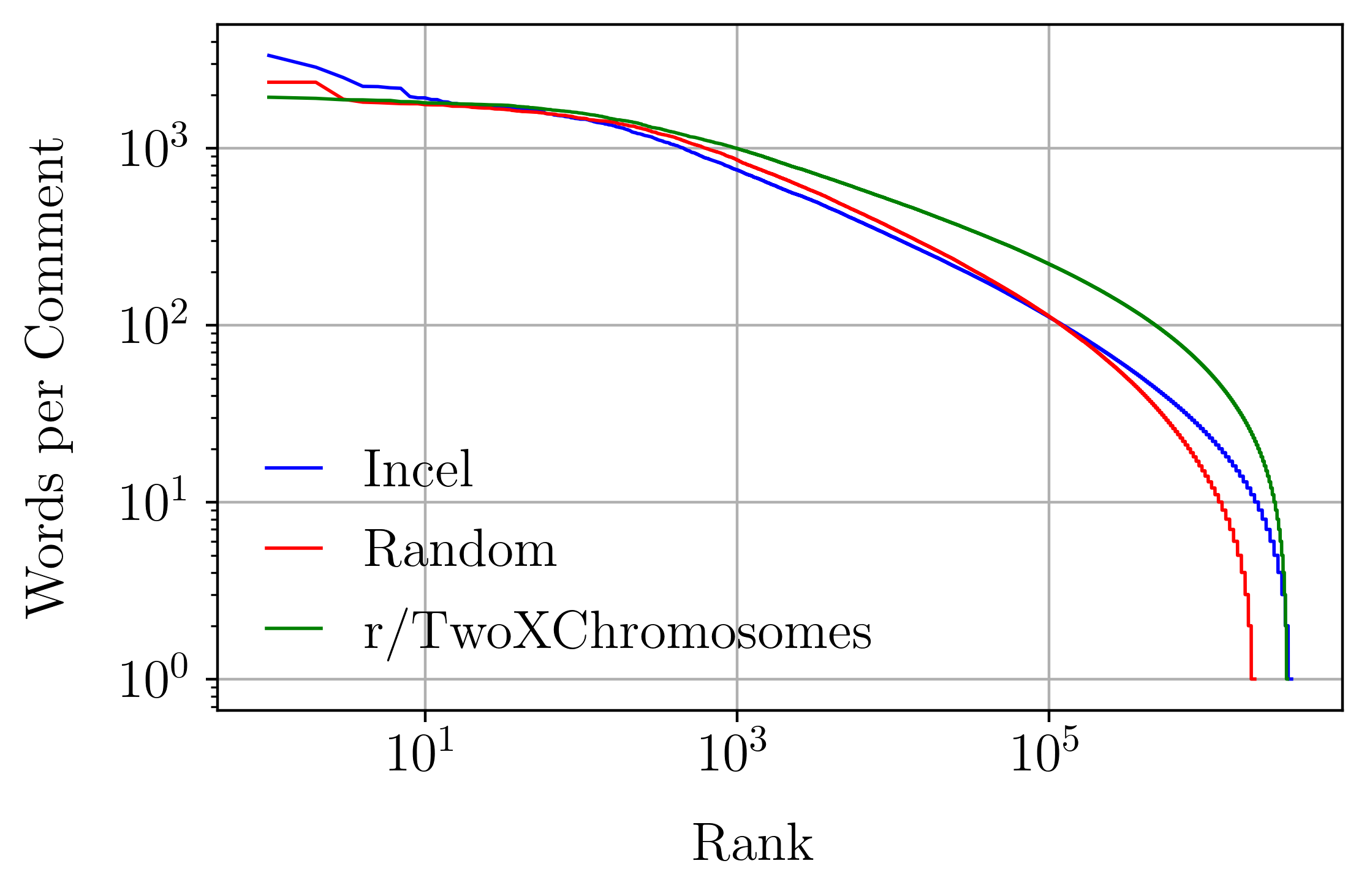}
\caption{\textbf{Zipf distribution of words per comment in the incel and random corpora.} The verbosity of each group of users could have an impact on the diversity of words used in comments.  To verify that the incel, random, and feminist users do not have vastly different comment lengths, we plot the Zipf distribution of the number of words in each comment. Results from pairwise Kolmogorov-Smirnov tests, corrected for large sample sizes, suggest that these distributions are not similar.  We also took 1,000 bootstrap samples from each distribution and computed the mean words per comment of each sample.  The distribution of mean words per comment for the feminist corpus is centered on 53.89, that of the random corpus is centered on 30.15, and that of the incel corpus is centered on 24.44.  Comments in the incel corpus are the least verbose out of all three corpora.}
\label{fig:zipf_word_comment1}
\end{figure}

We compare the word-rank distributions of the random and feminist corpora to the incel corpus.  The random corpus generation process differs from that of the incel corpus, and we acknowledge the differences in language that might be attributed to this.  The distribution of the total words per comment in Fig.~\ref{fig:zipf_word_comment1} is found to be not similar for the incel, random, and feminist corpora, according to pairwise Kolmogorov-Smirnov tests corrected for large sample sizes.  We took 1,000 bootstrap samples of size 10\% of each distribution and compared the distributions of mean words per comment for all three corpora.  We found that the the feminist corpus had the highest mean of the distribution of mean words per comment, 53.89.  The mean of the bootstrap distribution for the random corpus is 30.15, and for the incel corpus it is 24.44.  The incel comments tended to be the least verbose.

\begin{figure*}[t]
\centering
\includegraphics[width=\textwidth]{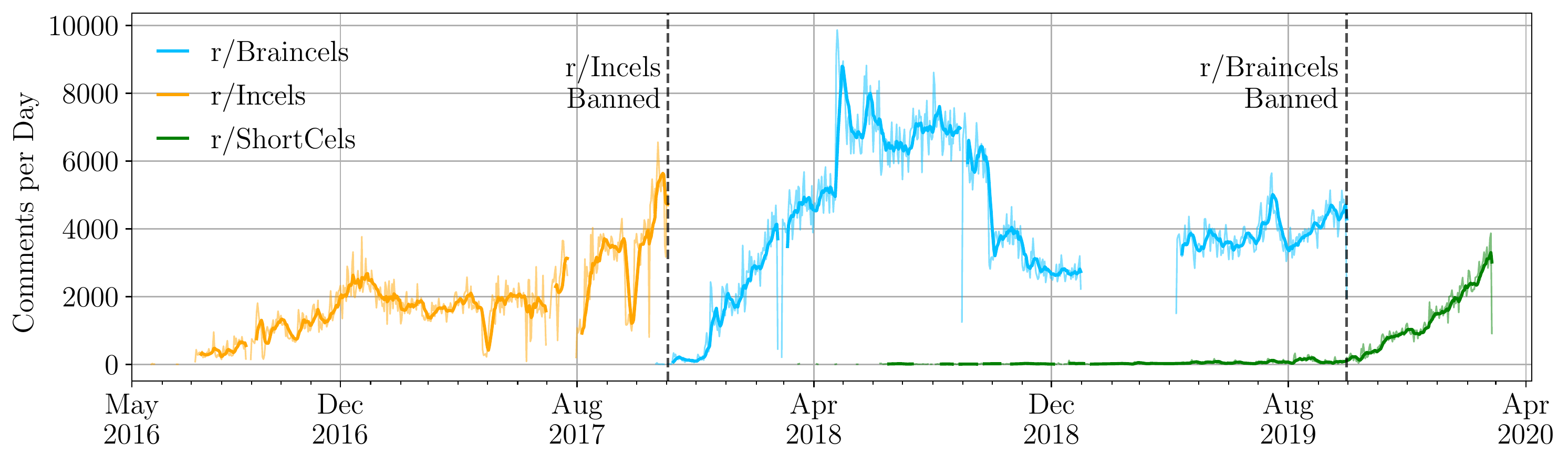}
\caption{\textbf{Total comments per day on r/Incels, r/Braincels and r/ShortCels}.  The timeseries above shows the count of comments per day on r/Incels, r/Braincels and r/ShortCels.  Each of these subreddits are considered to be central to incel activity on Reddit.  These timeseries support the theory that when an incel-associated subreddit is banned, another arises to take its place.  In addition to instituting bans, Reddit may issue quarantines on communities.  During quarantine, comments and posts can be published, but they are not visible to the rest of Reddit and were not included in the Pushshift Reddit dataset at the time.  Gaps in comment data represent periods of quarantine, such as the r/Braincels quarantine from December 31st, 2018 to April 6th, 2019.  The peak of comment activity in this timeseries is on the day of the Toronto van attack, a mass murder perpetrated in the name of inceldom.}
\label{fig:main_timeseries}
\end{figure*}


\subsection{Analysis}

We present a synoptic view of the comparison between the incel and random word-rank distributions by plotting the rank-rank pair of every word that occurs in the intersection of the incel and random corpora, and by ranking each term in descending order according to its contribution to the rank divergence between the two corpora \cite{dodds2020allotaxonometry}. We start by compiling the union of words, creating a combined lexicon containing the set of all words that occur in either corpus.  For each corpus, we rank words in the combined lexicon by each word's frequency.  The ranking is descending so that the most frequently occurring word has a rank of 1.  If $n$ words have the same frequency, the average of the next $n$ possible rankings is computed and each of those $n$ words share that average rank.  A rank-rank pair represents the rank of a single word in the incel corpus coupled with the rank of that word in the random corpus \cite{dodds2020allotaxonometry}.

To visualize the difference in rank between two corpora for all words in the combined lexicon, we present an allotaxonograph which includes a rotated histogram of logarithmic bins, colored by the density of rank-rank pairs contained in that bin (see Fig.~\ref{fig:rankdiv}, \ref{fig:women_bigrams} and \ref{fig:feminist_rankdiv}).  On the left side are words that had a higher rank in the random corpus, and on the right are words that had a higher rank in the incel corpus.  The more a word diverges from the center vertical line, the more heavily represented it is in the corresponding corpus.  The words that are labeled on the edges are randomly selected from the outermost bins to be the bin label (labels for the higher ranked, more dense bins are less meaningful), and these are words that have a notably higher rank in one corpus than the other, compared to the rest of the words.  This rank-rank histogram allows us to visualize the differences between corpora on both sides and at all levels of word frequency.  We are able to see function words at the top of the diamond, the richer words around the edges, and the least common words, or words that only exist in one corpus, at the bottom of the diamond \cite{dodds2020allotaxonometry}.

In addition to the holistic view provided by the rank-rank histogram, we compute rank-turbulence divergence, a measure of ``difference" between the incel and random corpora, and rank each terms by its contribution to divergence $D_{1/3}^{R}$ between these two systems, $\Omega_{incel}$ and $\Omega_{rand}$.  The absolute difference in the ranks of each word in the incel and random corpora, $r_{\tau,incel}$ and $r_{\tau,rand}$, are tuned by changing the $\alpha$ parameter, which dampens the importance of high frequency terms as it approaches 0. We choose $\alpha = 1/3$ in this case because it has been suggested for a moderate damping of high-frequency words \cite{dodds2020allotaxonometry}. Rank-turbulence divergence is the sum of each term's tunable absolute difference in inverted rankings between two corpora \cite{dodds2020allotaxonometry}.
\[ D_{1/3}^{R}(\Omega_{\textnormal{incel}} || \Omega_{\textnormal{rand}}) \propto \sum_{\tau} \left| \frac{1}{r_{\tau,\textnormal{incel}}^{1/3}} - \frac{1}{r_{\tau,\textnormal{rand}}^{1/3}} \right|^{3/4}\] 

In addition to the rank-rank histogram, the allotaxonograph lists the top 40 words, in descending order by divergence contribution, are listed in a rank shift plot.  These words are the most biased towards one corpus out of all words in the combined lexicon, and they are labeled by their rank-rank pairs.  Between the rank-rank histogram and the rank shift plot, the system balances are shown in three bar plots.  The top bar shows the percent of the total word count each system contains.  The middle bar shows the percent of the combined lexicon that each individual corpus contains, and the bottom bar shows the percentage of words that are exclusive to each corpus \cite{dodds2020allotaxonometry}.




We chose rank divergence as a measure of a word's imbalance between two corpora because it normalizes word frequencies by converting to rank and reduces the importance of high frequency words that may not be meaningful (e.g., function words such as `the' and `of'). Additionally, the rank-rank histogram and shift plot provide a clear way of visualizing the most key words in both corpora while maintaining quantitative dimensions like rank, density of words, and rank divergence contribution.

\section{Results and Discussion}\label{sec:results}
The subreddits we have chosen to study---r/Incels, r/Braincels, and r/Shortcels---appear to reflect a single community that respawns when its specific gathering site is banned. Fig.~\ref{fig:main_timeseries} supports this theory \cite{inceltimeline:2020,Futrelle.2019}.  The timeseries appear to be sequential, with notable activity surging when the previous channel is banned, which is indicative of ban evasion.  Ban evasion occurs when a subreddit is banned for violating Reddit's content policy and subsequently reorganizes into a new subreddit.  When identifying language patterns over time in this corpus of three different communities, we may encounter words that are unique to that particular community.  However, the words that are consistently present, ban evasion after ban evasion, give us confidence in a stable incel lexicon.

The Toronto van attack, an act of violence in the name of inceldom, precedes the most commented-on day in our dataset, April 24th, 2018.  The upward trend in comments per day on r/Braincels jumps up on this day, remaining above pre-attack levels until October 2018.  The rise in comment activity during and around this event may be due to increased incel activity, or due to increased traffic on incel subreddits because of media coverage of incels. The role played by media coverage in drawing attention to violent attacks is an important component of incel communication, but it is beyond the scope of the present work. 


Regarding the length of comments, incel channel contributors post more often than users in the random corpus and users in three other comparison groups. The incel comments data comes from specific forums, so the authors who comment in these forums are more likely to have a greater total number of comments, e.g., because they have an association to a specific forum.  Conversely, by randomly selecting comments from all of Reddit, it is unlikely that we capture many comments from a single author, because their comments are biased towards the subset of communities they participate in on Reddit.  This results in a bias towards fewer comments per author in the random corpus than in the incel corpus.  

\begin{figure}[t]
\centering
\includegraphics[width=\columnwidth]{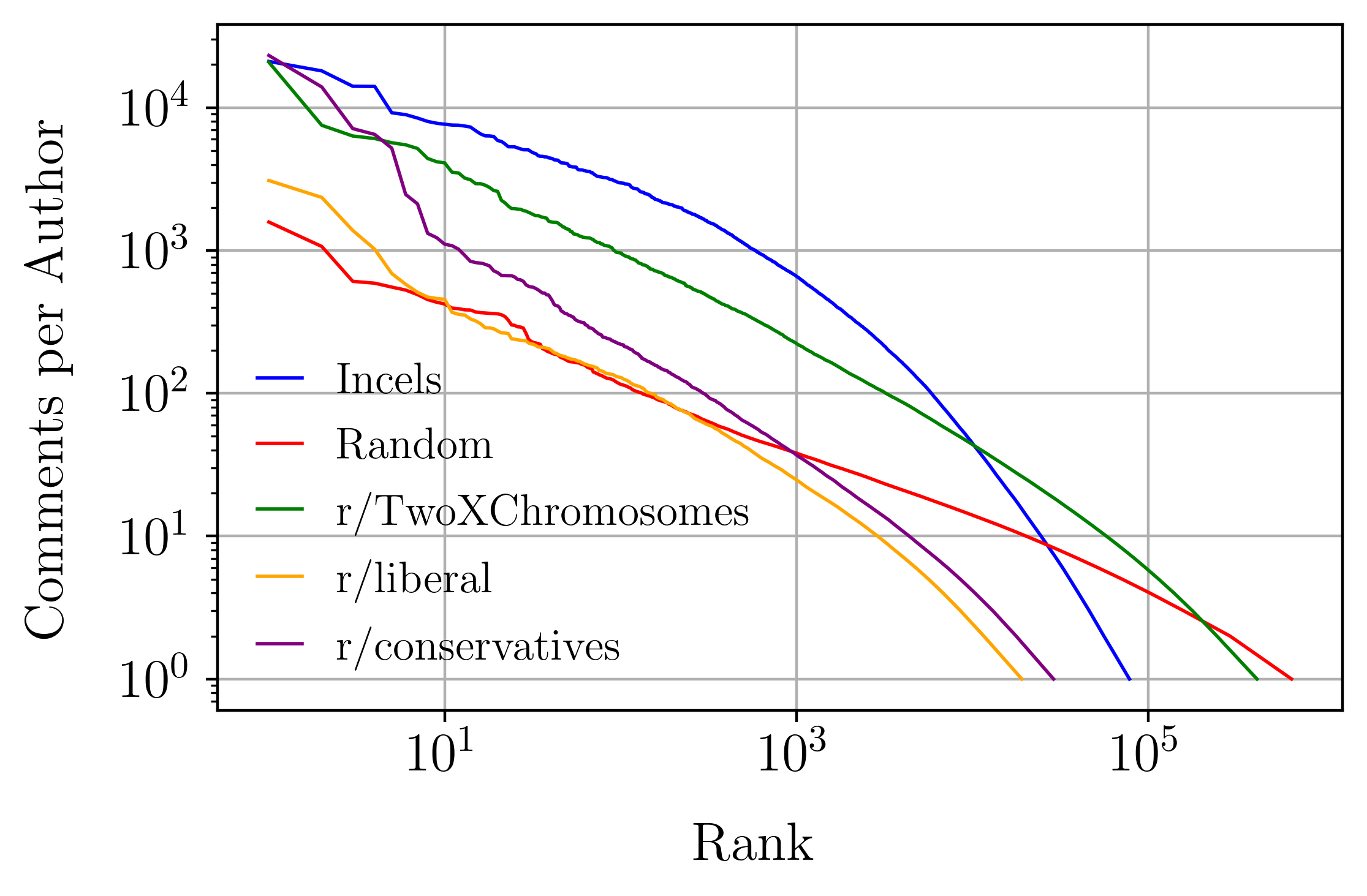}
\caption{\textbf{Zipf distribution of comments per author in the incel, random, r/TwoXChromosome, r/liberal, r/conservatives corpora.}  Incel users are more active and engaged on their respective subreddits than users on r/TwoXChromosome, r/liberal, r/conservatives, and users from randomly selected comments.  The Zipf distributions of comments per author for each corpus is shown to illustrate that the distribution for incel comments per author is greater than all other comparison corpora, which indicates a greater level of engagement from incel users.}
\label{fig:zipf_word_comment2}
\end{figure}

We also compare the number of comments per author in the incel corpus to that of r/TwoXChromosomes, r/liberal, and r/conservatives.  We chose r/liberal and r/conservatives as comparison groups because they represent common social ideologies to compare against the more extreme social ideology of inceldom.  We find that the distribution of comments per incel author indicates that incels comment more than all other comparison groups, see Fig.~\ref{fig:zipf_word_comment2}.

\begin{figure}[t]
\centering
\includegraphics[width=\columnwidth]{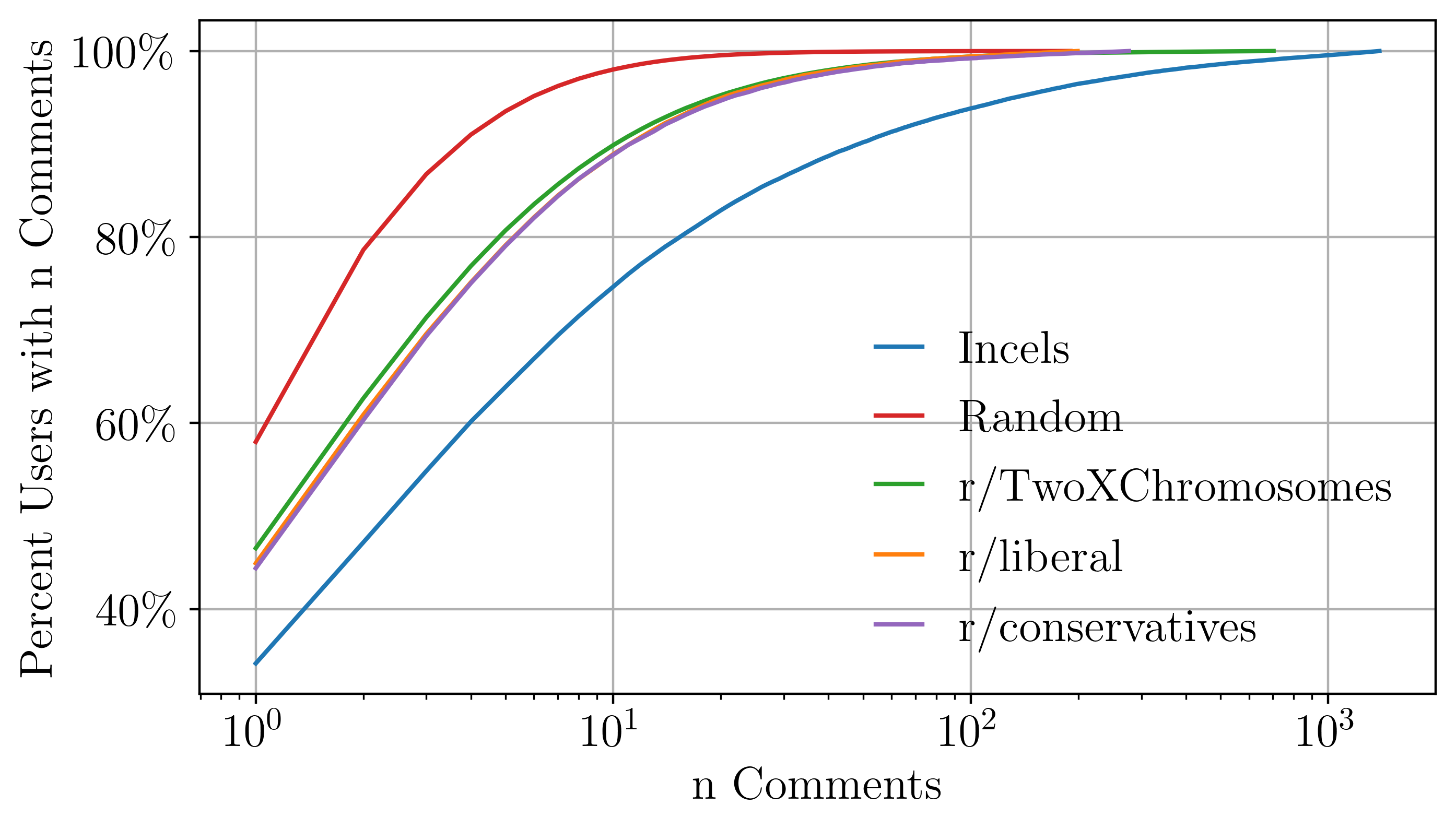}
\caption{\textbf{Complementary distribution function (CDF) of percentage of users from the random, incel, r/TwoXChromosomes, r/liberal, and r/conservatives corpora that commented n times.}  By plotting the CDF of the percentage of users that commented $n$ times for each corpus, we can identify that the group with the smallest fraction of single-comment users is the incel group.  The incel users are less likely to engage only once in their subreddit than users from our comparison communities.}
\label{fig:percent_once}
\end{figure}

Additionally, the percentage of users that only commented once is lowest in the incel corpus when compared to the random corpus, r/TwoXChromosomes, r/liberal, and r/conservatives corpora in Fig.~\ref{fig:percent_once}. These results suggest that incel community members are more engaged than users in other subreddits about social ideology.

\begin{figure}[t]
\centering
\includegraphics[width=\columnwidth]{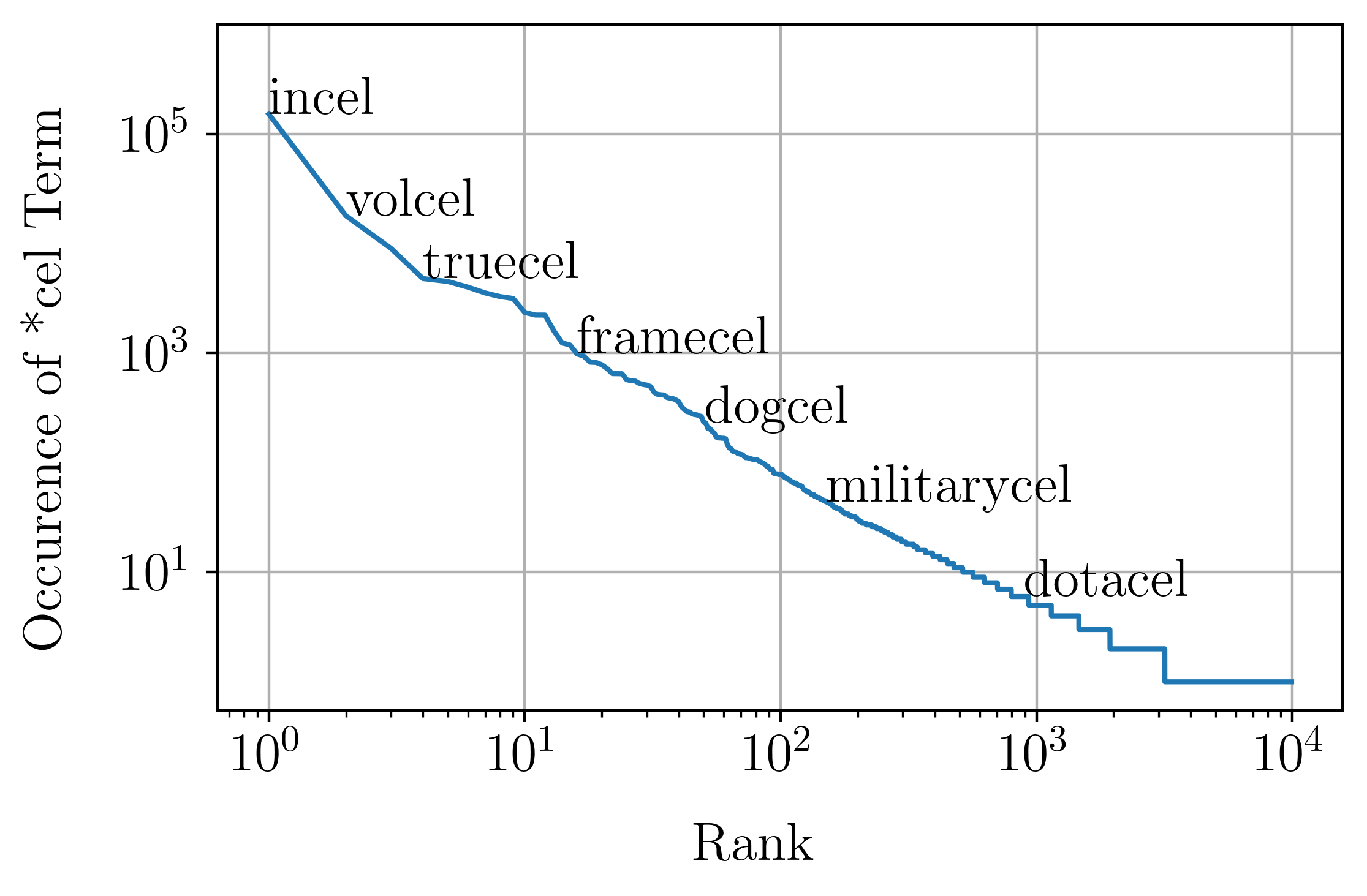}
\caption{\textbf{Zipf distribution of terms that end with ``-cel" in the incel corpus.} The above distribution shows the frequency of each term that ends in ``-cel" in the incel corpus vs its rank.  Some of the points have been labelled with their respective ``-cel" term.  The highest ranked and most frequently occurring term is ``incel".  ``Volcel", or ``voluntary celibate" is another popular instance of ``-cel" terms.  The diversity of these terms is indicative of user identification with the incel movement. }
\label{fig:cel}
\end{figure}

Terms that end in ``-cel" are popular among incel users who frequently create new instances of such terms to describe subtypes of incel.  Fig.~\ref{fig:cel} shows the annotated Zipf distribution of terms that end in ``-cel" in our corpus.  The most frequently used term is ``incel".  Other terms like ``volcel", which stands for ``voluntary celibate" are also used often.  More niche instances of ``-cel" terms include --- ``femcel" a female incel --- ``arabcel", an arab incel --- and ``dotacel", an incel who presumably plays the video game Dota.  Other popular instances are ``fakecel", someone who claims they are an incel but actually do have relationships with women, and ``truecel", the antithesis to the ``fakecel".  This type of term is meant to classify incels based on what makes them an incel, or some defining characteristic that sums up their identity.  Incels refer to themselves by unique ``-cel" terms to describe themselves and their life experience, resulting in a Zipfian distribution of frequency versus rank of ``-cel" terms.  Users on incel forums connect their own personal traits with the incel movement through language, making their own identity inseparable from the theoretical incel identity.

\begin{figure*}[t]
\centering
\includegraphics[width=\textwidth]{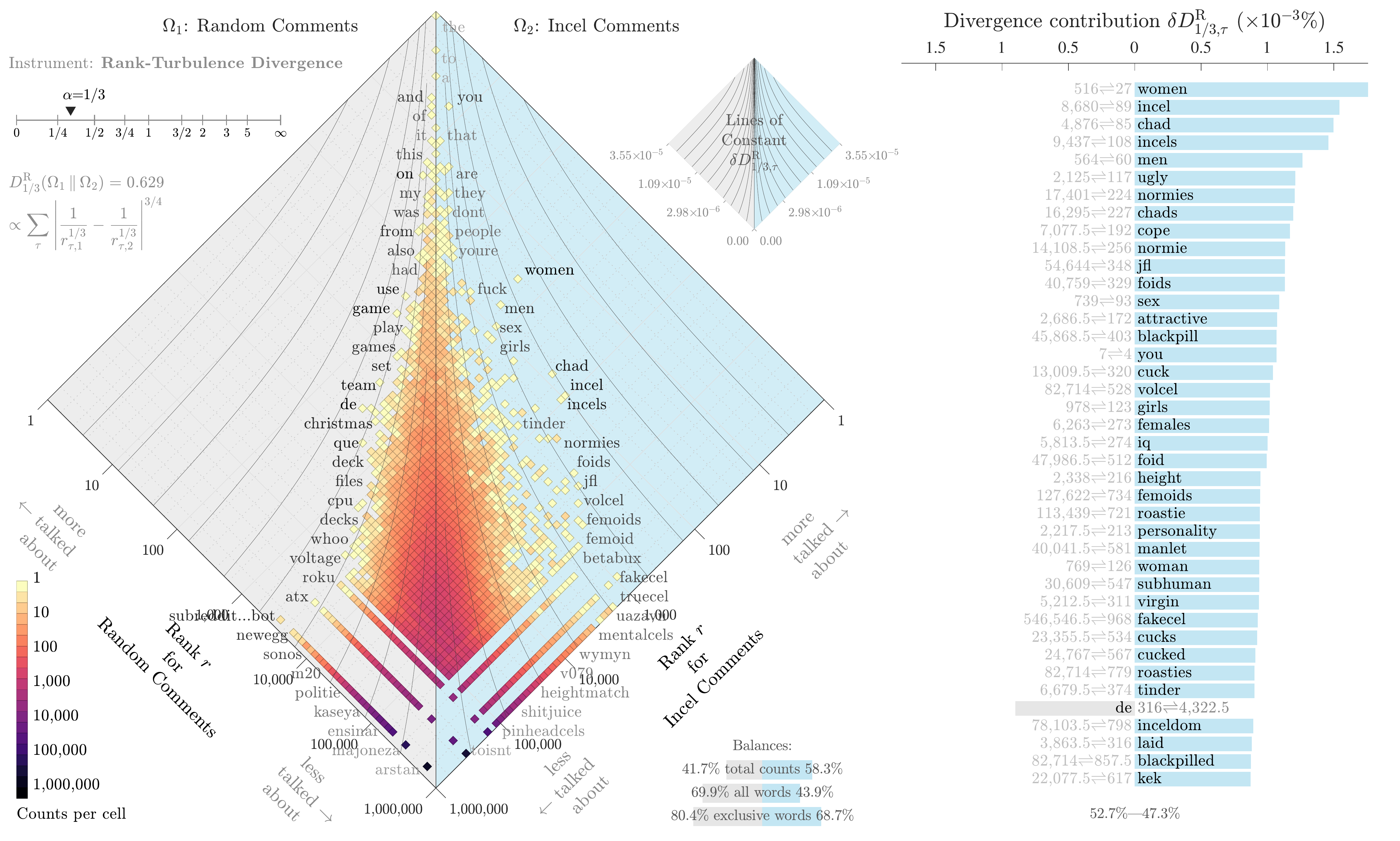}
\caption{\textbf{Rank-turbulence divergence allotaxonograph \cite{dodds2020allotaxonometry} of word rank distributions in the incel vs random comment corpora.} The rank-rank histogram on the left shows the density of words by their rank in the incel comments corpus against their rank in the random comments corpus.  Words at the top of the diamond are higher frequency, or lower rank.  For example, the word ``the" appears at the highest observed frequency, and thus has the lowest rank, 1.  This word has the lowest rank in both corpora, so its coordinates lie along the center vertical line in the plot.  Words such as ``women" diverge from the center line because their rank in the incel corpus is higher than in the random corpus.  The top 40 words with greatest divergence contribution are shown on the right.  In this comparison, nearly all of the top 40 words are more common in the incel corpus, so they point to the right.  The word that has the most notable change in rank from the random to incel corpus is ``women", the object of hatred and desire for the incel community.  The following words reference various categories of men: ``incels", ``chad", and ``men".  References to physical appearance are also more common in the incel corpus, such as ``ugly", ``attractive", and ``height".  A number of these words are made-up: ``normies", ``foids", ``blackpill", ``femoids", ``roastie", ``volcel", and ``fakecel".  These 40 words capture the real-life topics and made-up terms that populate the incel lexicon.}
\label{fig:rankdiv}
\end{figure*}

To construct our incel lexicon, we identify words that appear very often in incel comments and that also appear infrequently in random Reddit comments using rank-turbulence divergence.  Fig.~\ref{fig:rankdiv} is a rank-turbulence divergence plot of the distribution of words from our incel comments corpus compared to the distribution of words from our random comments corpus.  Nearly all of the top 40 words on the left, apart from ``you" and ``de", are thematically related to incel topics.  This indicates that while many words used in the random sample corpus are also used in the incel community, there are words distinctively used in the incel corpus, signaling the presence a distinct lexicon used by the incel community.

The set of words that appear far to the right of the center line are the words we hypothesize to be members of the incel lexicon.  The top 40 words that had the greatest difference in rank in the incel versus random corpora are listed in Fig.~\ref{fig:rankdiv}. Nearly every term is topically related to sex, gender, appearance, and social status, which indicates that the discussion in incel forums is topic-specific and homogeneous.
It is these words derived from rank-turbulence divergence that we identify as the empirical incel lexicon in Table~\ref{tab:lexicon}.  Our identified lexicon is comparable to online incel glossaries, written by and for the incel community \cite{incelglossary:2020}, but is revealed algorithmically in Fig.~\ref{fig:rankdiv}. As such, it confirms prior knowledge about the words that incels use to communicate with each other.  In the following analysis, we will study changes in the prevalence of incel language over time, and the contexts in which they are used.

\begin{table}[t]
    \centering
    \begin{tabular}{p{0.07\textwidth}
                p{0.16\textwidth}
                p{0.11\textwidth}
                p{0.16\textwidth}}
    \hline
         Term &  Random$\rightleftharpoons$Incel Rank & Term & Random$\rightleftharpoons$Incel Rank \\
    \hline
    & & & \\
        women & 516 $\rightleftharpoons$ 27 & foids & 40,759 $\rightleftharpoons$ 329\\
        incel & 8,680 $\rightleftharpoons$ 89 & sex &  739 $\rightleftharpoons$ 93\\
        incels & 9,437 $\rightleftharpoons$ 108 & attractive & 2,686.5 $\rightleftharpoons$ 172\\
        men & 564 $\rightleftharpoons$ 60 & blackpill & 45,868.5 $\rightleftharpoons$ 403\\
        ugly & 2,125 $\rightleftharpoons$ 117 & blackpilled &  82,714 $\rightleftharpoons$ 857.5\\
        normie & 14,108.5 $\rightleftharpoons$ 256 & femoids &  127,622 $\rightleftharpoons$ 734\\
        normies & 17,401 $\rightleftharpoons$ 224 & roastie & 113,439 $\rightleftharpoons$ 721\\
        chad & 4,876 $\rightleftharpoons$ 85 & roasties &  82,714 $\rightleftharpoons$ 779\\
        chads & 16,295 $\rightleftharpoons$ 227 & personality & 2,217.5 $\rightleftharpoons$ 213\\
        cope & 7,077.5 $\rightleftharpoons$ 192 & manlet &  40,041.5 $\rightleftharpoons$ 581\\
        foid & 47,986.5 $\rightleftharpoons$ 512 & virgin &  5,212.5 $\rightleftharpoons$ 311\\
    & & & \\
    \hline
    \end{tabular}
    \caption{\textbf{Empirical Incel Lexicon.} We have determined a lexicon of 22 terms that are more commonly used in comments on Incel subreddits than in randomly selected comments from all of Reddit.  For each term, we computed its rank-divergence contribution (see Fig.~\ref{fig:rankdiv}) to quantify the importance of each word's difference in its rank in the random corpus to its rank in the Incel corpus.}
    \label{tab:lexicon}
\end{table}

\begin{figure*}[t]
\centering
\includegraphics[width=\textwidth]{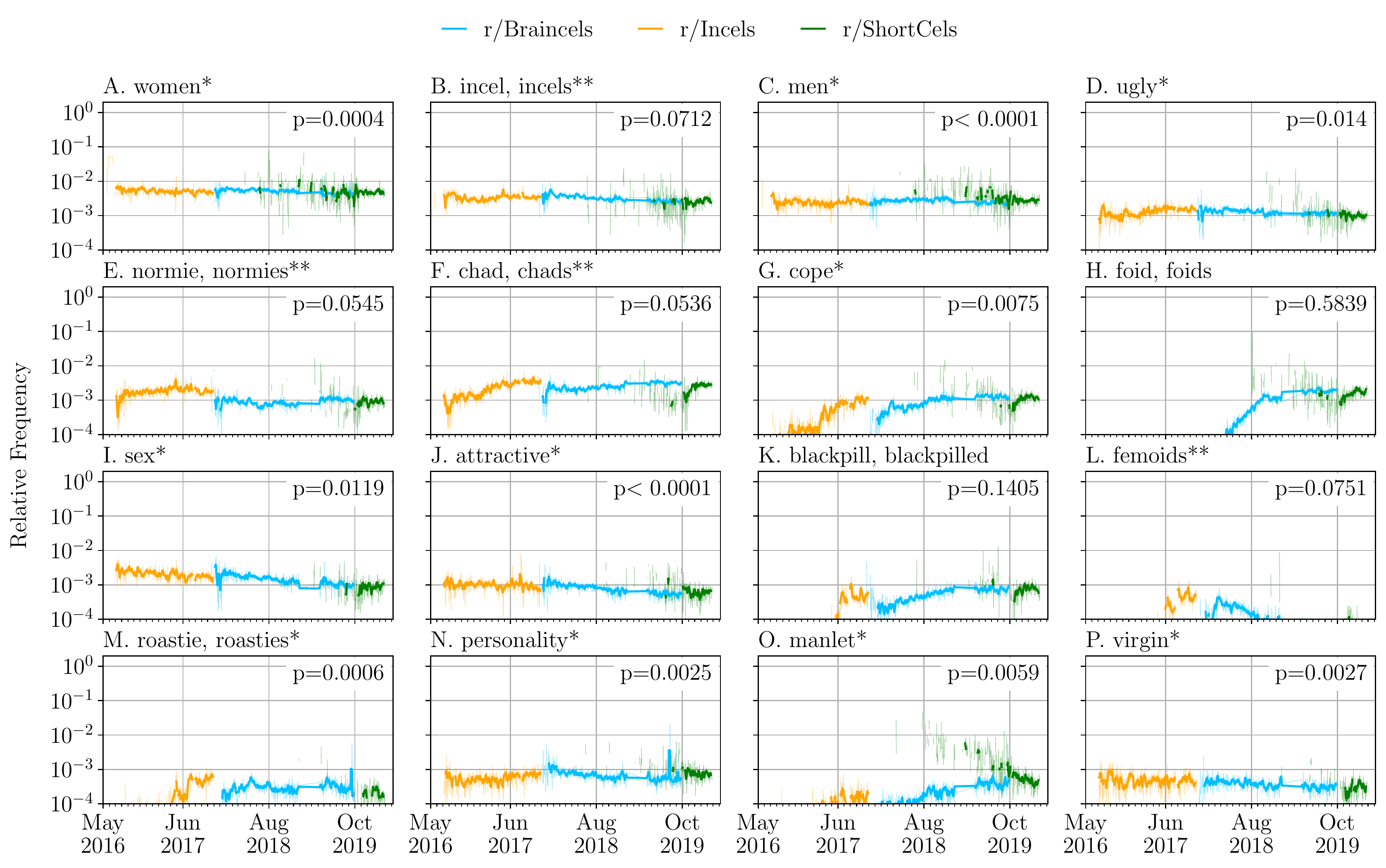}
\caption{\textbf{Timeseries of log relative frequency of empirical incel lexicon terms on r/Incels, r/Braincels, and r/ShortCels.} Each timeseries shows the ratio of total comments on a subreddit that contain a given word to total comments on that subreddit.  Here, we look at the timeseries of words that had high divergence contribution in Fig.~\ref{fig:rankdiv} on r/Incels, r/Braincels, and r/Shortcels.  In addition to the daily relative frequency values, a 14 day rolling average is shown.  We performed Augmented Dickey-Fuller tests for a unit root on each of these timeseries to test the null hypothesis that there is no unit root for these timeseries, or that they are stationary. ADF test for the timeseries in \textbf{A}, \textbf{C-D}, \textbf{G}, \textbf{I-J}, and \textbf{M-P} yielded a p-value below a significance threshold of 0.05, which means that these timeseries are stationary at the 0.05 significance level.  \textbf{B}, \textbf{E-F}, \textbf{L} are significant at the 0.1 significance level, therefore we are less confident that these timeseries do not have a unit root and are stationary.  Lastly, the ADF test could not show that \textbf{H} and \textbf{K} do not have unit roots and are stationary.  In \textbf{H}, the combined timeseries of ``foid" and ``foids" begins to increase in May 2018 and later stabilizes.  The timeseries in \textbf{K} for ``blackpill" and ``blackpilled" fluctuates between June and November 2017.}
\label{fig:word_timeseries}
\end{figure*}

To examine the stability of these words, we plot the timeseries of the incel lexicon in Fig.~\ref{fig:word_timeseries}.  A word may have had a short period of high frequency that contributed greatly to its rank, but may not reflect the incel lexicon as accurately as words that appear consistently over time.  The timeseries shown in Fig.~\ref{fig:word_timeseries} reveal that the relative frequency of the words ``women", ``men", ``incel", ``chad", ``cope", ``cuck", ``normies", ``virgin", and ``blackpill" are consistent over time, and consistent over three different communities.  To assess the stability of each of the timeseries in Fig.~\ref{fig:word_timeseries}, we perform Augmented Dickey-Fuller (ADF) tests for the unit root.  We reject the null hypothesis for the timeseries which yield a p-value below our significance threshold, 0.05.  These timeseries do not have a unit root, and are therefore stationary.  See Supplementary Table \ref{tab:dickey_fuller} for the ADF test summaries for each timeseries in Fig.~\ref{fig:word_timeseries}.

Consistent relative frequency over three banned communities shows us that the words in our identified lexicon may be a fingerprint for the larger incel community, not just for a single subreddit.  The bans of each community reduce the volume of comments per day, but the relative frequency of many of these words return to pre-ban levels.
The heavy bias of these words towards the incel corpus, along with their stable occurrences, give us a lexical fingerprint with which to identify the presence of incel communities in other corpora, and also give insight into the true values and interests of people who call themselves incels.  The root of their discussions is gender, with ``women" being the highest ranked incel term, ``men" ranked fifth, and a number of gender-based pejoratives ranked highly amongst ``men" and ``women".  This result supports the efficacy of our method for identifying the incel lexicon.

\begin{figure*}[t]
\centering
\includegraphics[width=\textwidth]{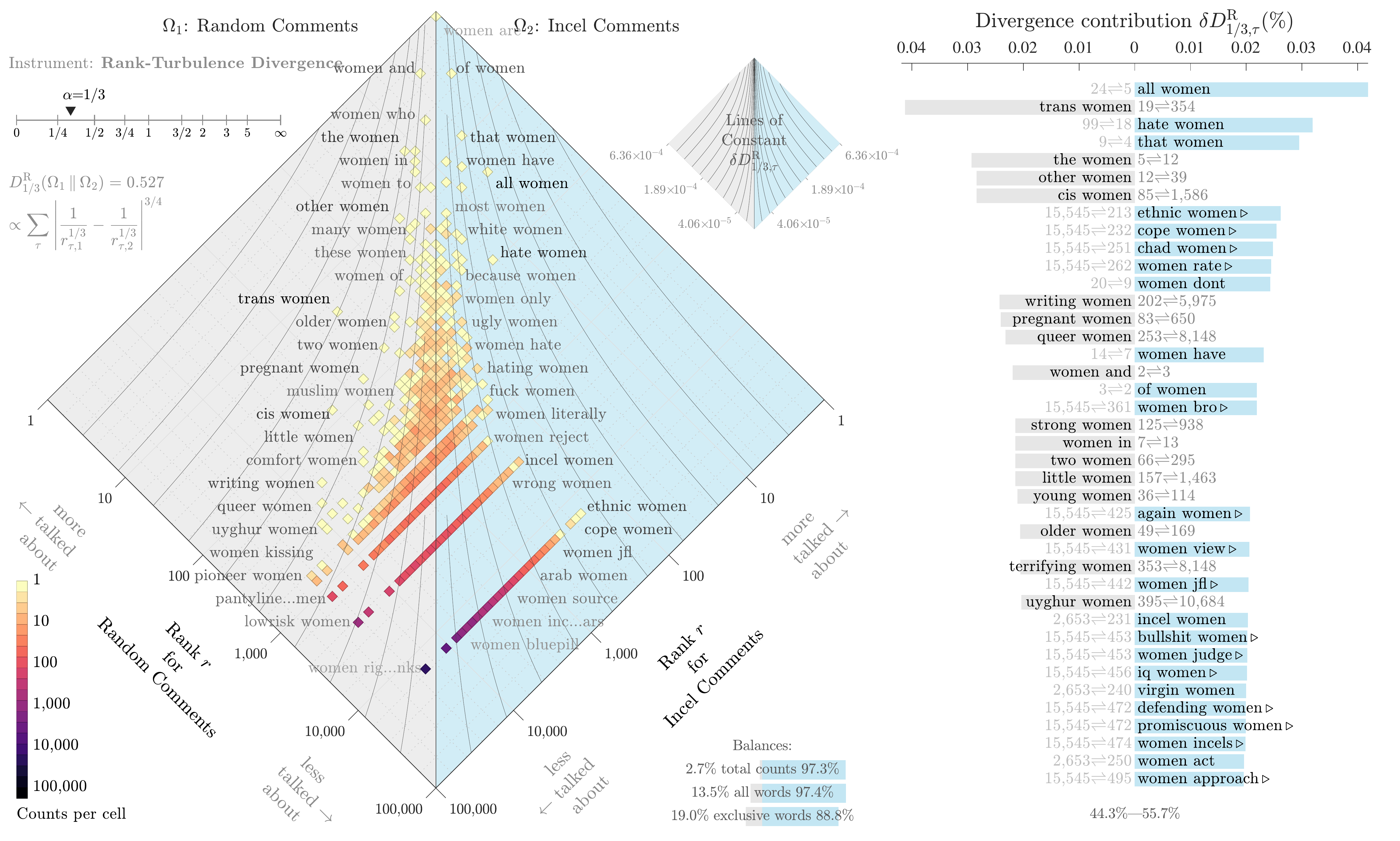}
\caption{\textbf{Rank-turbulence divergence allotaxonograph \cite{dodds2020allotaxonometry} of ``women" bigram distributions in the incel vs random comment corpora.} The top two bigrams that contain ``women" in the incel corpus are ``all women" and ``hate women".  The frequency of these bigrams matches prior assumptions that the incel community makes generalizations about women and view women with disdain.  Notably, female identities such as ``trans women" and ``pregnant women" are used far less in the incel corpus than in the random corpus.}
\label{fig:women_bigrams}
\end{figure*}

To contextualize the key incel terms that we have identified, we present common bigrams and trigrams that contain each term.  We focus on ``women" in Fig.~\ref{fig:women_bigrams}, and find that the bigrams with the highest positive difference in rank for the incel corpus fall into three main categories: generalizations about women, like ``all women"; hate speech, like ``hate women"; and modifications to other key incel terms like ``cope women", ``chad women", and ``incel women".  In the random corpus, female identities such as ``trans women", ``cis women", ``pregnant women", etc. are more common.  Fig.~\ref{fig:women_bigrams} suggests a limited and negative view towards women in Reddit incel communities.

We list the top 10 most incel-biased bigrams and trigrams in Table \ref{tab:bitrigrams} for ``women", ``men", ``ugly", and ``virgin".  Top 10 bigrams and trigrams for all other words in the identified lexicon are listed in Supplementary Table \ref{tab:sup_bitrigrams}.  Trigrams of key incel terms can give us new information that bigrams cannot.  For example, bigrams containing ``women" reference generalizations about women, various verbs that ``women" is associated with, and identities such as ``ethnic women" or ``chad women".  The trigrams that contain ``women", however, are far more indicative of hatred towards women, and reference ``all women" more often.  We do not give a discrete list of topics associated with each word, but rather a gallery of common word co-occurrences that encompass the broad and diverse context around each key incel term.  Other popular incel themes are represented in this list of bigrams and trigrams, such as the Pareto principle, or the ``80/20 rule".  Incels often claim that, assuming men and women are ranked by attractiveness, the top 80\% of women seek relationships with the top 20\% of men, leaving the bottom 80\% of men to seek relationships with the bottom 20\% of women.  This principle is often referred to as the foundation for why incels cannot successfully create romantic relationships and is rooted in the assumption that women are hypergamous, or that they only seek relationships with men who are more attractive than them.

\begin{table*}[h]
\begin{center}
\begin{tabular}{ c c c }
  \hline
  \thead{1-gram} & \thead{Bigrams} & \thead{Trigrams} \\
  \hline
    women &  \makecell{\\all women\\
hate women\\
that women\\
ethnic women\\
cope women\\
chad women\\
women rate\\
women dont\\
women have\\
of women\\\\}  & \makecell{\\all women are\\
i hate women\\
hate women i\\
women i hate\\
women are the \\
not all women\\
hate women so\\
that all women\\
women dont care\\
women are evil\\\\}  \\
  \hline
  ugly & 
  \makecell{\\ugly men\\
ugly people\\
ugly guys\\
are ugly\\
ugly male\\
an ugly\\
ugly guy\\
ugly short\\
youre ugly\\
ugly males\\\\}  & 
\makecell{\\if youre ugly\\
you are ugly\\
an ugly guy\\
an ugly man\\
for being ugly\\
an ugly male\\
an ugly face\\
short and ugly\\
he is ugly\\
i am ugly\\\\}  \\
\end{tabular}
\begin{tabular}{ c c c }
  \hline
  \thead{1-gram} & \thead{Bigrams} & \thead{Trigrams} \\
  \hline
  men &  \makecell{\\ugly men\\
short men\\
men are\\
virgin men\\
of men\\
attractive men\\
sub8 men\\
ethnic men\\
men fuck\\
men have\\\\}  & \makecell{\\of men are\\
20 of men\\
80 of men\\
percent of men\\
men are the\\
short men are\\
good looking men\\
majority of men\\
ugly men are\\
of ugly men\\\\}  \\
  \hline
  virgin & 
  \makecell{\\virgin shaming\\
virgin men\\
kissless virgin\\
non virgin\\
old virgin\\
virgin girl\\
virgin women\\
virgin that\\
ugly virgin\\
lonely virgin\\\\}  & 
\makecell{\\being a virgin\\
a kissless virgin\\
be a virgin\\
die a virgin\\
a virgin so\\
are a virgin\\
wants a virgin\\
isnt a virgin\\
a virgin than\\
virgin shaming is\\\\}  \\
\end{tabular}
\begin{tabular}{ c c c }
  \hline
    incel(s) &  \makecell{\\be incel\\
not incel\\
youre incel\\
every incel\\
incel but\\
im incel\\
are incel\\
all incels\\
still incel\\
incels here\\\\}  & \makecell{\\not an incel\\
as an incel\\
an incel but\\
of an incel\\
be an incel\\
cant be incel\\
all incels are\\
that incels are\\
an incel you\\
you are incel\\\\}  \\
  \hline
    normie(s) &  \makecell{\\normies are\\
failed normie\\
normies will\\
failed normies\\
normie and\\
by normie\\
all normies\\
tier normies\\
normie here\\
normie tier\\\\}  & \makecell{\\high tier normie\\
im a normie\\
a normie and\\
a failed normie\\
fuck off normie\\
youre a normie\\
a normie i\\
not a normie\\
normies and chads\\
low tier normie\\\\}  \\
  \hline
\end{tabular}
\begin{tabular}{ c c c }
  \hline
    chad &  \makecell{\\fuck chad\\
if chad\\
chads and\\
chad doesnt\\
want chad\\
chad cock\\
chad can\\
chad and\\
white chads\\
chads cock\\\\}  & \makecell{\\not a chad\\
a chad and\\
with a chad\\
was a chad\\
a chad or\\
chad and stacies\\
to be chad\\
only want chad\\
a chad is\\
of a chad\\\\}  \\
  \hline
    sex & 
  \makecell{\\get sex\\
have sex\\
pity sex\\
getting sex\\
deserve sex\\
owed sex\\
sex isnt\\
gets sex\\
women sex\\
meaningless sex\\\\}  & 
\makecell{\\entitled to sex\\
can get sex\\
sex with chad\\
sex sex sex\\
cant have sex\\
have sex with\\
cant get sex\\
not having sex\\
to get sex\\
sex with us\\\\}  \\
\hline
\end{tabular}
\caption{\textbf{Top 10 rank-divergence contributing bigrams and trigrams containing incel terms.} In a closer look at eight terms from our identified incel lexicon, we have determined the top 10 bigrams and trigrams with the greatest rank-divergence contribution that contain a term from the lexicon.  By comparing the rank distributions of bigrams containing ``women" in the incel corpus versus that distribution in the random corpus, we are able to identify the bigrams and trigrams that are most biased towards the incel corpus and contribute the most to the overall rank-divergence between the incel and random distributions.}
\label{tab:bitrigrams}
\end{center}
\end{table*}

In addition to comparing the incel corpus to the random corpus, we compare the incel corpus to a feminist corpus, comments from r/TwoXChromosomes.  This comparison gives us information on how language used by members of the ideological opposite of inceldom, feminism, differ from their misogynistic counterparts.  The top 40 rank-divergence contributing words in the incel-feminist comparison are shown in the rank shift plot below in Fig.~\ref{fig:feminist_rankdiv}.  This list contains many of the same words that appear in the top 40 words from the incel-random corpus comparison, as well as some additional profanity.  Note that ``women" has a higher rank in the incel corpus than it does in this feminist corpus, an indicator of how deeply engaged incel users are in discussing women over all other topics.

\begin{figure*}[t]
\centering
\includegraphics[width=\textwidth]{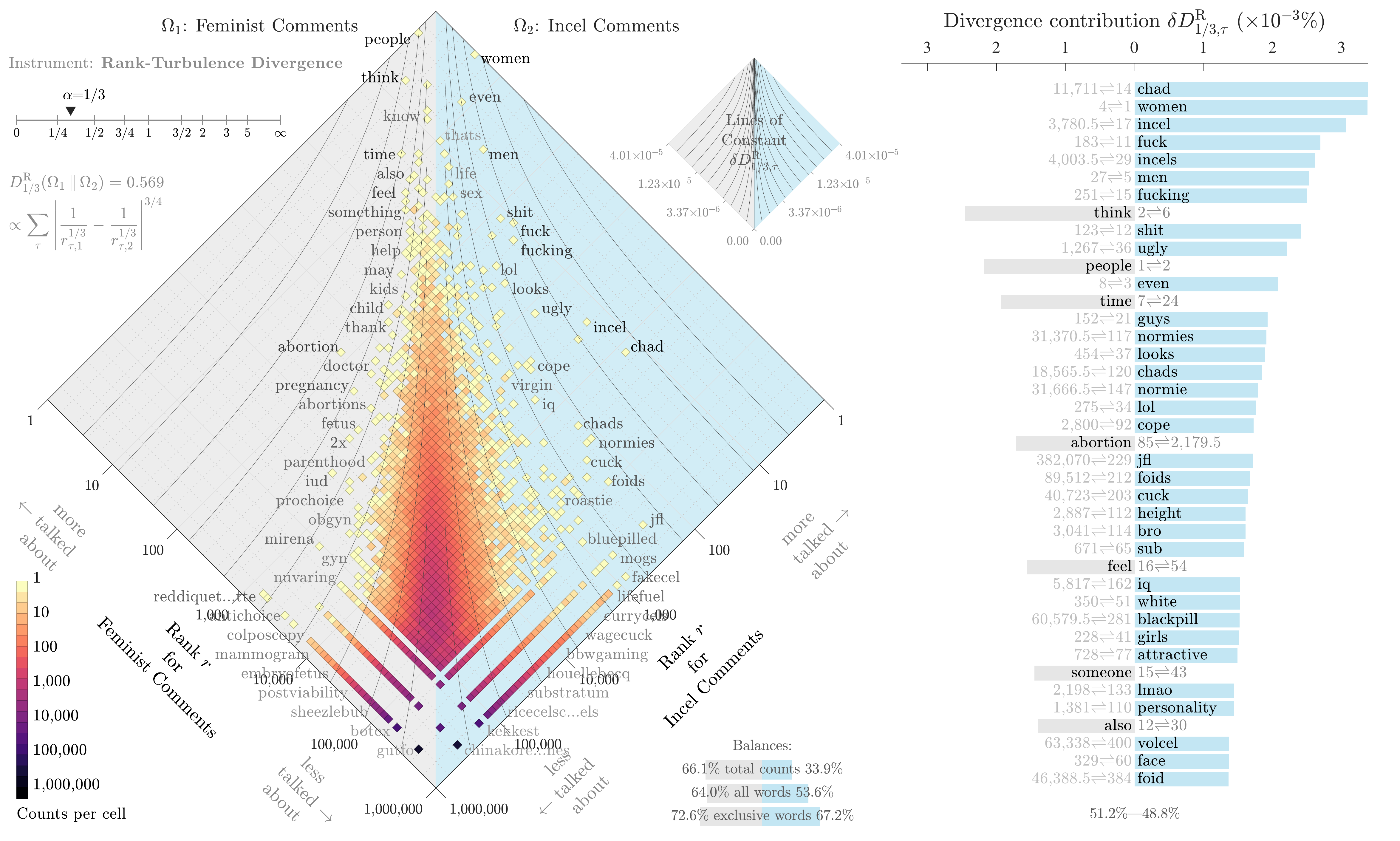}
\caption{\textbf{Rank-turbulence divergence allotaxonograph \cite{dodds2020allotaxonometry} of word rank distributions in the incel vs feminist comment corpora.} This allotaxonograph compares the word rank distribution of the incel comment corpus to that of the feminist comment corpus by plotting rank-rank pairs in the histogram on the left and by listing the top 40 rank-divergence contributing words to the right.  Many of the top 40 words in this comparison also appear in the incel-random corpus comparison in Fig.~\ref{fig:rankdiv}.  It is notable that ``women", the second term in the list of top 40 rank-divergence contributing words, has a higher rank in the incel corpus than in the feminist corpus.  This result suggests that conversation on incel forums is even more centered around women than on feminist forums.}
\label{fig:feminist_rankdiv}
\end{figure*}

We have empirically derived an incel lexicon and explored the contextual meaning behind incel terminology.  To gain an understanding of the dominance of these terms in the incel lexicon over time, we examine the evolution of the most ``narratively dominant" word by comparing corpora of incel comments from a single month to comments from 12 months prior, and listing the word with the greatest increase in rank in the most recent month (see Table \ref{tab:nar_dom}) \cite{dodds2020computational}. Results for this analysis with 2 month and 6 month gaps are recording in Supplementary Tables \ref{tab:two_month} and \ref{tab:six_month}.  The terms ``chad", ``men", ``foids", ``short", and ``height" remain dominant for more than one month, suggesting a regime of popularity for each of these words during their respective months of narrative dominance. 
 
\section{Concluding Remarks}\label{sec:conclusion}
Previous data-driven studies on incel-related communities have aimed to identify key words, to make inferences about incel demographics, to characterize activity on various platforms, and to automate detection of related communities \cite{jaki2019online, farrell2019exploring, ribeiro2020pick, papadamou2020understanding}.  A study by Jaki \textit{et. al.} used statistical tests on word counts to identify key words in incel corpora, and visualized their results as a word cloud \cite{jaki2019online}.  Our approach differs from the approach used by Jaki \textit{et. al.} because we use a principled statistical method,
rank-turbulence divergence with a rank-rank histogram and rank shift plot, to visualize words associated with incel comments.  Farrel \textit{et. al.} create various lexicons to categorize themes on incel forums, but we do not utilize this method as it may introduce bias from the researchers, as well as bias towards particular topics due to imbalance in the size of each topic's lexicon \cite{farrell2019exploring}.  In other work by Farrell \textit{et. al.}, they use word embeddings to identify a unique misogynist lexicon using Reddit post data from popular misogynist subreddits \cite{10.1145/3394231.3397912}.  Additionally, differences in sentiment of comments from male-dominated versus female dominated subreddits were computed and modeled using BERT and beta regression models \cite{aggarwal2020exploration}.  Our work does not compute sentiment and does not use machine learning models to compare corpora.  

We contribute to research on online incel communities by using rank-turbulence divergence as a measure for comparing the word rank distributions of these communities to a random comments set.  This method reduces the bias that is introduced by comparing word counts alone, as well as the bias introduced by spurious results from statistical testing when performed on large data sets. Additionally, we avoid limitations brought on by predefined lexicons by identifying words that are important to incel communication based on their frequency of appearance.  Lastly, rank-turbulence divergence is interpretable and can be visualized such that the rank of the word in each corpus, as well as its rank-divergence contribution, can be understood within the context of other words in the corpora.

Violence against women by misogynists persists through events like the Toronto van attack and more subtly through language on online forums like those we examined. Algorithmic detection and moderation of these forums remains a challenge due to the decentralized nature of the incel movement, anonymity online, and ever changing language patterns.  Future work on the incel community could use incel language patterns to detect the presence of incel and incel-like communities.  Additionally, groups like Men Going Their Own Way and Pick Up Artists hold many of the same views and values as incels, and should be studied to better understand misogyny in online social networks.

The Reddit incel community, known for its pessimistic and misogynistic views, has revealed a lexicon that, when compared to the rest of Reddit, is clearly identifiable.  By creating a glossary of bigrams and trigrams containing terms in the incel lexicon, we provide a look into the diverse contexts in which these terms are used.  The discussion surrounding the most incel-biased term, ``women", is largely hateful and confirms our prior assumption that the incel community is toxic and misogynistic.  Additionally, we gain insight into how incels view themselves.  For example, the words that occur with ``ugly" refer most often to men.  By identifying an incel lexicon and the context that surrounds the words in the lexicon, we summarize key terms and topics that appear on incel forums.

\acknowledgments 
The authors are grateful for the computing resources provided by the Vermont Advanced Computing Core which was supported in part by NSF award No. OAC- 1827314 and financial support from the Massachusetts Mutual Life Insurance Company. 

\bibliography{references}

\clearpage

\newwrite\tempfile
\immediate\openout\tempfile=startsupp.txt
\immediate\write\tempfile{\thepage}
\immediate\closeout\tempfile

\setcounter{page}{1}
\renewcommand{\thepage}{S\arabic{page}}
\renewcommand{\thefigure}{S\arabic{figure}}
\renewcommand{\thetable}{S\arabic{table}}
\setcounter{figure}{0}
\setcounter{table}{0}

\section{Supplementary Information (SI)}




\onecolumngrid\
\begin{table}[!ht]
    \centering
    \begin{tabular}{c|c|c}
    \hline
        Word(s) & ADF Statistic & p-value \\
    \hline
        women* & -4.340 & 0.0004 \\
        incel, incels** & -2.717 & 0.0712 \\
        men* & -5.786 & $< 0.0001$ \\
        ugly* & -3.32 & 0.014 \\
        normie, normies** & -2.827 & 0.0545 \\
        chad, chads** & -2.834 & 0.0536 \\
        cope* & -3.519 & 0.0075 \\
        foid, foids & -1.397 & 0.5839 \\
        sex* & -3.373 & 0.0119 \\
        attractive* & -4.864 & $< 0.0001$ \\
        blackpill, blackpilled & -2.405 & 0.1405 \\
        femoids** & -2.694 & 0.0751 \\
        roastie, roasties* & -4.218 & 0.0006 \\
        personality* & -3.842 & 0.0025 \\
        manlet* & -3.595 & 0.0059 \\
        virgin* & -3.817 & 0.0027 \\
    \hline
    \end{tabular}
    \caption{Augmented Dickey-Fuller (ADF) tests for timeseries of relative term frequency in empirical incel lexicon.  For each term in our empirical incel lexicon, we performed an ADF test where our null hypothesis is that there is no unit root in the timeseries of the relative frequency of a word, and is therefore stationary. * p-value is significant at the 5\% critical value.  ** p-value is significant at the 10\% critical value.}
\label{tab:dickey_fuller}
\end{table}
\twocolumngrid\

\begin{table*}[!th]
\begin{center}
\begin{tabular}{>{\centering}p{0.05\textwidth}
                >{\centering}p{0.21\textwidth}
                p{0.21\textwidth}}
  \hline
  \thead{1-gram} & \thead{Bigrams} & \thead{Trigrams} \\
  \hline
femoids &  \makecell{\\femoids are\\
femoids and\\
of femoids\\
and femoids\\
femoids have\\
to femoids\\
that femoids\\
femoids dont\\
femoids will\\
femoids would\\\\}  & \makecell{\\all femoids are\\
femoids are the\\
normies and femoids\\
femoids are not\\
that femoids are\\
femoids would rather\\
femoids have no\\
i hate femoids\\
all the femoids\\
femoids are so\\\\}  \\
  \hline
  attractive & 
  \makecell{\\attractive men\\
attractive people\\
attractive than\\
attractive personality\\
facially attractive\\
by attractive\\
attractive trait\\
consider attractive\\
attractive chad\\
attractive dudes\\\\}  & 
\makecell{\\attractive to women\\
less attractive than\\
are more attractive\\
an attractive face\\
finds you attractive\\
attractive than me\\
an attractive guy\\
attractive men are\\
they find attractive\\
if youre attractive\\\\}  \\
\end{tabular}
\begin{tabular}{>{\centering}p{0.05\textwidth}
                >{\centering}p{0.21\textwidth}
                p{0.21\textwidth}}
  \hline
  \thead{1-gram} & \thead{Bigrams} & \thead{Trigrams} \\
  \hline
 foid & 
  \makecell{\\to foids\\
and foids\\
foids will\\
foids have\\
foids in\\
that foids\\
foids would\\
this foid\\
foid who\\
foid i\\\\}  & 
\makecell{\\to a foid\\
with a foid\\
for a foid\\
foids are the\\
a foid and\\
a foid who\\
of a foid\\
all foids are\\
all the foids\\
a foid i\\\\}  \\
  \hline
    roastie(s) &  \makecell{\\roasties are\\
a roastie\\
the roasties\\
and roasties\\
roastie is\\
fuck roasties\\
roasties fuck\\
of roasties\\
the roastie\\
roastie i\\\\}  & \makecell{\\fuck roasteies fuck\\
roasties fuck roasties\\
used up roastie\\
fuck off roastie\\
post wall roastie\\
normies and roasties\\
is a roastie\\
a roastie and\\
a roastie is\\
with a roastie\\\\}  \\
\end{tabular}
\begin{tabular}{>{\centering}p{0.05\textwidth}
                >{\centering}p{0.21\textwidth}
                p{0.21\textwidth}}
  \hline
 personality & 
  \makecell{\\personality its\\
good personality\\
personality is\\
bad personality\\
about personality\\
personality detector\\
personality bro\\
your personality\\
personality matters\\
personality meme\\\\}  & 
\makecell{\\your personality its\\
a good personality\\
its your personality\\
personality its your\\
a bad personality\\
per...ty doesnt matter\\
a shit personality\\
personality is the\\
shit about personality\\
care about personality\\\\}  \\
  \hline
    cope & 
 \makecell{\\a cope\\
is cope\\
cope im\\
cope you\\
massive cope\\
biggest cope\\
it cope\\
cope no\\
its cope\\
cope but\\\\}  & 
\makecell{\\is a cope\\
cope if you\\
a good cope\\
its a cope\\
the biggest cope\\
the best cope\\
as a cope\\
just a cope\\
this is cope\\
cope cope cope\\\\}  \\
  \hline
\end{tabular}
\begin{tabular}{>{\centering}p{0.05\textwidth}
                >{\centering}p{0.21\textwidth}
                p{0.21\textwidth}}
  \hline
 manlet & 
  \makecell{\\the manlet\\
manlet with\\
manlet in\\
manlet but\\
ugly manlet\\
manlet and\\
turbo manlet\\
manlet is\\
manlet so\\
ethnic manlet\\\\}  & 
\makecell{\\im a manlet\\
being a manlet\\
youre a manlet\\
is a manlet\\
as a manlet\\
manlet with a\\
was a manlet\\
a manlet is\\
a manlet with\\
a manlet but\\\\}  \\
  \hline
   blackpill(ed) & 
  \makecell{\\blackpill i\\
brutal blackpill\\
blackpilled and\\
a blackpilled\\
are blackpilled\\
to blackpill\\
atomic blackpill\\
blackpill the\\
being blackpilled\\
blackpill to\\\\}  & 
\makecell{\\the blackpill i\\
with the blackpill\\
blackpill is the\\
take the blackpill\\
the atomic blackpill\\
spread the blackpill\\
the blackpill but\\
are some blackpill\\
some blackpill truths\\
to the blackpill\\\\}  \\
\hline
\end{tabular}
\caption{\textbf{Incel lexicon bigrams and trigrams: Top 10 incel-biased bigrams and trigrams containing terms from the lexicon.} We determined the top 10 bigrams and trigrams that contain a term from the lexicon.  By comparing the rank distributions of bigrams containing a particular term in the incel corpus versus that distribution in the random corpus, we are able to identify the bigrams and trigrams that are most biased towards the incel corpus.}
\label{tab:sup_bitrigrams}
\end{center}
\end{table*}

\onecolumngrid\
\begin{table}[!th]
\begin{center}
\begin{tabular}{>{\centering}p{0.19\textwidth}
                >{\centering}p{0.19\textwidth}
                >{\centering}p{0.19\textwidth}
                >{\centering}p{0.19\textwidth}
                p{0.19\textwidth}}
  \hline
  \thead{Month} &
  \thead{2016-2017} & \thead{2017-2018} & \thead{2018-2019} & \thead{2019-2020} \\
  \hline
  Jan  &  & incels  & short  & \multicolumn{1}{c}{fuck} \\
  Feb  &  & men  &  short & \multicolumn{1}{c}{women} \\
  Mar  &  & men  & short  &  \\
  Apr  &  & men  & honk  &  \\
  May  & fuck & men  & chad  &  \\
  Jun  & one & men & chad &  \\
  Jul  & chad & foids  &  height &  \\
  Aug  & chad & foids  & fuck  &  \\
  Sep & chad & foids  & ha  &  \\
  Oct  & chad & foids  & short  &  \\
  Nov  & chad & foids &  height &  \\
  Dec  & women & even  & height  &  \\
  \hline
\end{tabular}
\caption{\textbf{Month vs 12 Previous Months Narratively Dominant Terms.} The term that had the greatest increase in rank from a particular month to that month, one year later, is recorded to identify patterns in monthly narrative dominance.  Each row in this table corresponds to the most recent compared month, and each column describes the years that were compared for a given month.}
\label{tab:nar_dom}
\end{center}
\end{table}
\twocolumngrid\

\begin{table*}[!th]
\begin{center}
\begin{tabular}{>{\centering}p{0.16\textwidth}
                >{\centering}p{0.16\textwidth}
                >{\centering}p{0.16\textwidth}
                >{\centering}p{0.16\textwidth}
                >{\centering}p{0.16\textwidth}
                p{0.15\textwidth}}
  \hline
  \thead{Month} &
  \thead{2016} & \thead{2017} & \thead{2018} & \thead{2019} & \thead{2020} \\
  \hline
  Jan-Feb  &  & chad  & men  & tallfags & \multicolumn{1}{c}{peoplei} \\
  Feb-Mar  &  & chad  &  even & gun  & \\
  Mar-Apr  &  & sex  & headsup  & women  & \\
  Apr-May  &  & get  & fuck  &  chad & \\
  May-Jun  &  & men  & get  &  get & \\
  Jun-Jul  &  & blackops2cel & people & women &  \\
  Jul-Aug  & people & units  &  get & personality &\\
  Aug-Sep  & women & fuck  & bagel  &  ha & \\
  Sep-Oct & one & refresh  & even  &  short & \\
  Oct-Nov  & men & sex  & irs  &  get & \\
  Nov-Dec  & people & see &  chad & koku & \\
  Dec-Jan  & women & think  & short  & white &  \\
  \hline
\end{tabular}
\end{center}
\label{tab:two_month}
\caption{\textbf{Month vs one previous month narratively dominant terms.}  The term that had the greatest increase in rank from a particular month to that month, one year later, is recorded to identify patterns in monthly narrative dominance.  Each row in this table corresponds to the most recent compared month, and each column describes the years that were compared for a given month.}
\end{table*}

\begin{table*}[!th]
\begin{center}
\begin{tabular}{>{\centering}p{0.19\textwidth}
                >{\centering}p{0.19\textwidth}
                >{\centering}p{0.19\textwidth}
                >{\centering}p{0.19\textwidth}
                p{0.19\textwidth}}
  \hline
  \thead{Month} &
  \thead{2016-2017} & \thead{2017-2018} & \thead{2018-2019} & \thead{2019-2020} \\
  \hline
  Jan-Jul  &  & chad  & even  & \multicolumn{1}{c}{fuck} \\
  Feb-Aug  &  & chad  & foids  & \multicolumn{1}{c}{women}   \\
  Mar-Sep  &  & men  & white  & \multicolumn{1}{c}{even}  \\
  Apr-Oct  &  & fuck  & foids  & \multicolumn{1}{c}{short}  \\
  May-Nov  & sex & sex  & fucking  & \multicolumn{1}{c}{height}   \\
  Jun-Dec  & think & incels & even & \multicolumn{1}{c}{height}   \\
  Jul-Jan  & 314 & incels  & short & \multicolumn{1}{c}{short}\\
  Aug-Feb  & women & incels & short & \multicolumn{1}{c}{peoplei}  \\
  Sep-Mar  & chad & incels  & short &   \\
  Oct-Apr  & chad & men & honk  &   \\
  Nov-May  & chad & men & chad &   \\
  Dec-Jun  & chad & men & white  &   \\
  \hline
\end{tabular}
\caption{\textbf{Month vs six previous months narratively dominant terms.}  The term that had the greatest increase in rank from a particular month to that month, one year later, is recorded to identify patterns in monthly narrative dominance.  Each row in this table corresponds to the most recent compared month, and each column describes the years that were compared for a given month.}
\end{center}
\label{tab:six_month}
\end{table*}

\end{document}